\documentclass[superscriptaddress,showpacs,twocolumn,prl,floatfix]{revtex4-1}
\usepackage{amsfonts}
\usepackage{amsmath}
\usepackage{amssymb}
\usepackage{graphicx}
\usepackage{epstopdf}

\begin{document}

\title{Spin edge helices in a perpendicular magnetic field} 
\author{S. M. Badalyan}
\email{Samvel.Badalyan@physik.uni-regensburg.de}
\affiliation{Department of Physics, University of Regensburg, 93040 Regensburg, Germany}
\affiliation{Department of Radiophysics, Yerevan State University, 1 A. Manoukian St., Yerevan, 375025 Armenia}
\author{J. Fabian}
\affiliation{Department of Physics, University of Regensburg, 93040 Regensburg, Germany}
\date{\today}

\begin{abstract}
We present an exact solution to the problem of the spin edge states in the presence of 
equal Bychkov-Rashba and Dresselhaus spin-orbit fields in a two-dimensional electron system, 
restricted by a hard-wall confining potential and exposed to a perpendicular magnetic field. 
We find that the spectrum of the spin edge states depends critically on the orientation of the
sample edges with respect to the crystallographic axes. Such a strikingly different 
spectral behavior generates new modes of the persistent spin helix---spin edge helices 
with novel properties, which can be tuned by the applied electric and magnetic fields.
\end{abstract}

\pacs{72.25.-b, 72.15. Gd, 85.75.-d}
\maketitle

\paragraph{Introduction.}

Edge states are a defining factor in many prominent transport phenomena in condensed matter physics. They emerge and are protected against scattering in the quantum Hall systems by applying a perpendicular magnetic field to a two-dimensional conductor, or are formed at the interfaces of topological insulators  \cite{Hasan2010:P} by deforming the bulk band structure in the presence of strong spin-orbit coupling. The interplay of the three effects: the magnetic field quantization, the spin-orbit coupling, and the confinement by sample edges, is yet largely unexplored. 

Spin-orbit coupling is an important tool to manipulate electron spins in solids by purely electric means~\cite{as, zfds,fmesz}. In particular,  the spin Hall effect~\cite{dyakonov} or the spin Hall drag~\cite{shd} allow to create spin accumulation across the transport channels. The Bychkov-Rashba (BR) spin-orbit coupling~\cite{br}, due to the structure inversion asymmetry, and the Dresselhaus (D) coupling~\cite{d}, due to the confinement quantization of the bulk spin-orbit interaction (SOI) dominate in many semiconductor heterostructures.  Both BR and D couplings can be tuned by electric gates, asymmetric doping, or strain, allowing for efficient spin control~\cite{schliemann,stano,bavf2009,bavf2010}. If the strengths of the BR and D couplings are equal, there can be long-lived, persistent spin helices (PSH) formed and protected against spin relaxation~\cite{bernevig}. This exciting phenomenon was recently observed~\cite{koralek,fabian}.

Here we study the complex interplay of the SOI, cyclotron effects of an external magnetic field, and the hard wall confinement in a generic zinc-blende two-dimensional electron system (2DES) grown along [001]. 
We present an exact solution to the problem of the spin edge states for the equal BR and D SOI strengths. The spectrum of the edge states is strikingly different if the edges run along $[110]$ or $[1\overline{1}0]$ directions. Depending on the relative sign of the BR and D couplings, in those two symmetry directions either (a) the spectrum is spin degenerate or (b) the spin splitting of the edge channels becomes maximal, i.e. either spin polarized channels and spin current oscillations~\cite{smb2009} are possible or not. We find that in (a) the non-Abelian gauge field via the built-in spin-dependent phase factor generates spin edge helices (SEH) with a precession angle that depends on the transverse distance from the edge. In (b) the shifting property of the spectrum allows the existence of SEH with a precession angle that depends on the distance along the propagation direction. In strong magnetic fields the precession angle of the SEH is quantized in case (a),  while a periodic helical structure, extended along the edge, is produced in case (b). In weak magnetic fields we find interesting new spin resonances when the cyclotron motion is commensurate with the spin precession~\cite{rashbasheka}. Experimentally, a strong reduction of spin scattering rate towards the sample edges should be observed. 

As an important application of our theory we propose extending the experimental setup in Ref.~\onlinecite{koralek} by exposing the 2DES additionally to a perpendicular magnetic field $B$. According to the experimental findings from  Ref.~\onlinecite{koralek}, the enhancement of the PSH by about two-order of magnitude occurs only at intermediate temperatures about $T\sim 100$ K. Its rapid drop with lowering of $T$ shows that the spin Coulomb drag~\cite{ida,weber,smb} via a strong increase of the diffusion coefficient destroys the PSH enhancement, for which a finite momentum $Q$ is needed in $B=0$. In the case of finite $B$ we find that in order to excite SEH in the (b) configuration a finite momentum shift is still needed between the spin components but in the (a) configuration, where the energy is degenerate, the SEH exists for $Q=0$, i.e. without momentum difference along its propagation direction. This should reduce the role of spin Coulomb drag in suppressing the enhancement of SEH, which can be useful for spintronic applications. Notice that in three-dimension a nonequivalence in $(1\bar{1}0)$ and $(\bar{1}10)$ has been already observed by means of the electron paramagnetic resonance and the electron-dipole spin resonance~\cite{rashbasheka,dobrowolska}.

\paragraph{Theoretical concept.}

\begin{figure}[t]
\includegraphics[width=0.47\linewidth]{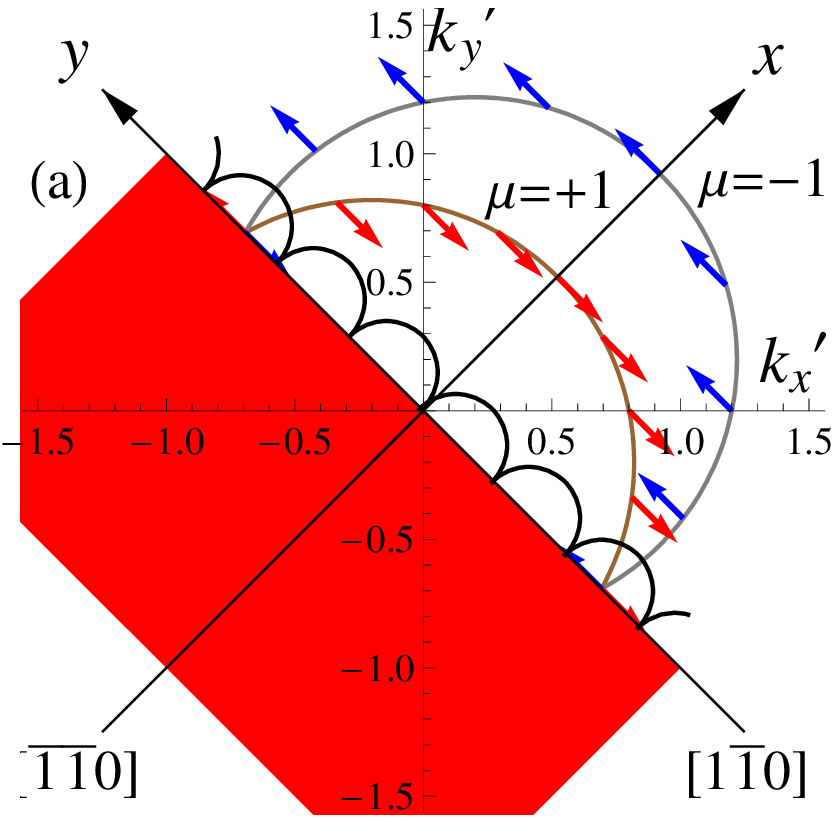} \hspace{0.03\linewidth} %
\includegraphics[width=0.47\linewidth]{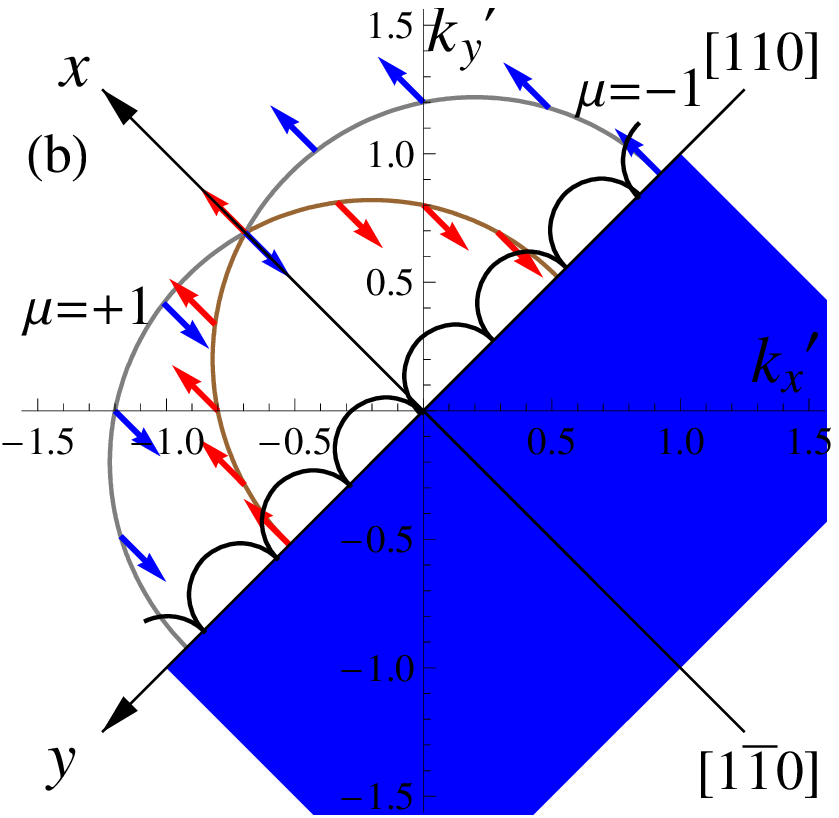} \hspace{0.03\linewidth} %
\caption{(Color online) Two physically different configurations of the 2DES, restricted with a hard-wall confining potential  at $x=0$ (filled areas). The spin edge states propagate in the skipping orbits along $[1\bar{1}0]$ and $[110]$ axes, respectively, parallel and perpendicular to the direction of electron spins. The Fermi contours in the absence of the magnetic field are shown in the momentum plane with arrows indicating the directions of spin.}
\label{config}
\end{figure}

We consider electrons in a 2DES 
exposed to a perpendicular magnetic field $B$ along [001]. The electrons are additionally confined by an infinite potential $V\left( x\right) =\infty $ for $x<0$. Then in the presence of the BR and D SOI the Hamiltonian is $H=H_{0}+H_{SOI}+V(x)$
where $H_{0}=\vec{\pi}^{2}/2m^{\ast }$ desribes a free particle in a quantizing magnetic field, $m^{\ast }$ denotes the electron effective mass and $\vec{\pi}=\vec{p}-(e/c)\vec{A}$ the kinetic momentum with $\vec{p}=-i\hbar \vec{\nabla}$; the electron charge is $-e$. There are two preferential directions in the $\left( x,y\right) $ plane of 2DES: i) the direction of the sample boundary along which edge states propagate, and ii) the direction of the electron spin in the presence of BR and D SOI of equal strength, determined by the crystallographic axes. The relative orientation of these two directions determines two distinct configurations, shown in Fig.~\ref{config}. In these configurations we choose the $(x,y)$ coordinate system such that the sample boundary is always along $y$. 
Then  in the configuration (a) in Fig.~ \ref{config}, in which the axes $x$ and $y$ are along $[110]$ and $[\overline{1}10]$, we have
\begin{equation}
H_{SOI}=\left( \alpha _{R}-\alpha _{D}\right) \pi _{y}\hat{\sigma}_{x}-\left( \alpha _{R}+\alpha _{D}\right) \pi
_{x}\hat{\sigma}_{y}~.
\end{equation}%
In the configuration (b), the axes $x$ and $y$ are along $[\overline{1}10]$ and $[\overline{1}\overline{1}0]$ and the Hamiltonian is
\begin{equation}
H_{SOI}=\left( \alpha _{R}+\alpha _{D}\right) \pi _{y}\hat{\sigma}_{x}-\left( \alpha _{R}-\alpha _{D}\right) \pi _{x}\hat{\sigma}_{y}~.
\end{equation}
Here $\alpha _{R}$ and $\alpha _{D}$ are, respectively, BR and D spin-orbit coupling constants and
 $\hat{\sigma}_{x},\hat{\sigma}_{y} $ are the Pauli matrices. 
We consider magnetic fields which are strong enough to quantize the electron spectrum but weak enough to cause much smaller Zeeman splitting than the SOI induced energy splitting~\cite{kramer2005,smb2009}.

We choose the Landau gauge with $\vec{A}\left( x\right) =\left( 0,xB,0\right)$ and making the ansatz 
\begin{equation}
\psi \left(\vec{r}\right) =e^{ik_{y}y}\chi _{k_{y}}(x)~,~ \vec{r}=(x,y)~,  \label{eq5}
\end{equation}%
reduce the two-dimensional Schr\"{o}dinger equation to the effective one-dimensional problem in the $x$ direction. 
The transformed Hamiltonian $U^{\dag }HU$, generated by the global unitary transformations 
\begin{equation}
U^{a}=\frac{1}{\sqrt{2}}\left(\begin{array}{cc} i & -i \\  1 & 1\end{array} \right)~\text{and}~ 
U^{b}=\frac{1}{\sqrt{2}}\left( \begin{array}{cc} -1 & 1 \\ 1 & 1\end{array}\right)~,
\label{unitary}
\end{equation}
becomes diagonal in the (a) and (b) configurations so that in the case of $\alpha _{R}=\alpha _{D}$=$\alpha $ the wave functions $\chi_{k_{y}}\left( x\right)$ satisfy the following equation
\begin{equation}
\left[ \left( \frac{d}{dx}+i\gamma a \hat{\sigma}_{z} \right) ^{2}+\mu +\frac{1}{2}-\frac{\left( x-X\left( k_{y}\right) 
-\gamma b \hat{\sigma}_{z} \right) ^{2}}{4}\right] \chi_{k_{y}}\left( x\right) =0~.
\label{shreq}
\end{equation}
Here the coefficients $a$ and $b$ are given by 
\begin{gather}
\begin{array}{ccc}
a=1, & b=0, & \text{for the (a) configuration}\end{array}~, \\
\begin{array}{ccc}
a=0, & b=2, & \text{for the (b) configuration}\end{array}~.
\end{gather}%
In Eq.~\ref{shreq} $\mu=\nu +\gamma ^{2}$ and we express the total electron energy $E\rightarrow \left( \nu +1/2\right)\hbar \omega _{B}$ and the length $x\rightarrow xl_{B}/\sqrt{2}$ in units of the cyclotron energy, $\hbar \omega_{B}=\hbar eB/m^{\ast }c$, and the magnetic length, $l_{B}\equiv \sqrt{\hbar c/e B}$. The dimensionless SOI coupling constant $\gamma =\sqrt{2}\alpha /v_{B}$, where the cyclotron velocity $v_{B}=\hbar /m^{\ast }l_{B}$, while the dimensionless coordinate of the center of orbital rotation $X\left( k_{y}\right) =\sqrt{2}k_{y}l_{B}$.

In the (a) configuration the SOI with equal BR and D strengths induces a finite non-Abelian gauge field that depends on the spin orientation, keeping the cyclotron rotation center unshifted. In contrast, in the (b) configuration the SOI induces only a spin-dependent shift of the cyclotron rotation center. 

With the unitary transformation 
\begin{equation}
\chi_{k_{y}}\left( x\right) =\exp \left( -i\gamma a x \hat{\sigma}_{z}\right) \phi_{k_{y}}\left( x\right) ~,
\end{equation}
we can eliminate the non-Abelian gauge field and map the SOI problem of the equal BR and D strengths to the SOI free, shifted edge state problem as follows:
\begin{equation}
h_{\mu }\left( x-s\gamma b\right) \phi^{s} _{k_{y}}\left( x\right) =0~.
\label{eq11}
\end{equation}%
Here we introduce the operator
\begin{equation}
h_{\nu }\left( x\right) =\left( \frac{d^{2}}{dx^{2}}+\nu +\frac{1}{2}-\frac{\left( x-X\left( k_{y}\right) \right) ^{2}}{4}\right) ~  
\label{eq8}
\end{equation}
and $s=\pm 1$ labels the spin $\uparrow $ and $\downarrow $ states in the new spin basis, created by (\ref{unitary}). 
The general solution of Eq.~\ref{eq11} is given in terms of the parabolic cylindric functions $D_{\mu }\left( x\right)$ so that the spin edge states are given by
\begin{equation}
\psi _{k_{y}}^{s}\left( \vec{r} \right) =\exp \left( -i s\gamma a x +i k_y y \right) 
D_{\mu}\left( x-X\left( k_{y}\right) -s\gamma b\right) ~.  
\label{eq12}
\end{equation}%
For sufficiently large positive $X\left( k_{y}\right) $ the solution (\ref{eq12}) recovers the exact spectrum of the dispersionless bulk Landau levels, $E_{sl}(\gamma )=\left( l+\frac{1}{2}-\gamma ^{2}\right) \hbar \omega _{B}$
with the index $l=0,1,2\ldots $~\cite{wang}; this is valid in both configurations since the index $\mu $ is the same. The shift of all bulk Landau levels due to SOI is independent of spin so they remain spin degenerate.  

\paragraph{Energy spectrum.}

\begin{figure}[t]
\includegraphics[width=0.75\linewidth]{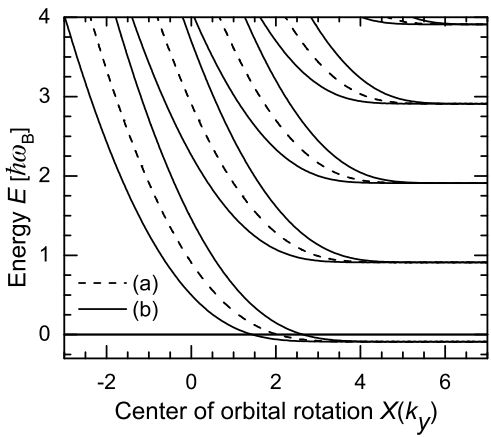}
\caption{The energy spectrum of spin edge states in the presence of Bychkov-Rashba and Dresselhaus 
SOI of equal strengths. The dashed and solid curves correspond to the (a) and (b) configurations in Fig.~1.}
\label{spectrum}
\end{figure}

The energy spectrum of the spin edge states is obtained by letting the wave functions (\ref{eq12}) vanish at the sample boundary: $\left. D_{\mu \left( \nu ,\gamma \right) }\left( x-X(k_{y})-s\gamma b\right)\right\vert _{x=0}=0$.
The calculated energy branches, $E_{sl}(k_{y},\gamma )$, are shown in Fig.~\ref{spectrum}.  As seen in the (a) configuration the energy spectrum remains spin degenerate for all values of $k_{y}$ so that the spin edge states at the Fermi energy, $E_{F}$, are not separated in space. The wave functions (\ref{eq12}) contain finite spin-dependent phase factors ($a=1$), which describe the precession of spins with the opposite rotations along the electron propagation axis $y$.

In contrast, the energy spectrum in (b) develops, for a given principal quantum number $l$, two spin branches. The degeneracy is lifted already for positive $X\left( k_{y}\right) \sim 2$; the spin-splitting increases strongly with $|X\left( k_{y}\right) |$ for negative $X\left( k_{y}\right) $. The spin-dependent phase factor in (\ref{eq12}) vanishes identically ($a=0$) and the effect of equal BR and D SOI on the spin edge states is reduced to the spin-dependent shift of the energy branches in the momentum space. At certain values of $E_F$ there exist two spatially separated current carrying states with opposite spins, whose directions are locked globally by the geometry of the configuration (b).

Thus, the spectrum in Fig.~\ref{spectrum} shows that the external magnetic field creates edge states along the $[1\bar{1}0]$ and $[110]$. The linear motion of the edge states via the BR and D SOI of equal strengths induces an effective magnetic field along $[110]$ and $[1\bar{1}0]$, respectively, that is either perpendicular or parallel to the direction of the spins thereby keeping the spin degeneracy or resulting in the maximal spin-splitting, respectively, in (a) or (b). 

\paragraph{Spin edge helices.}
The energy bands of the spin edge states possess a shifting property along $y$, 
\begin{equation}
E_{\downarrow l}(k_{y},\gamma )=E_{\uparrow l}\left( k_{y}-Q,\gamma \right),  
\label{eq15}
\end{equation}
where the shifting wavenumber $Q=\sqrt{2}\gamma b/l_{B}$ is finite in (b) and zero in (a). As in the zero field~\cite{bernevig}, we introduce the following operators in the transformed spin basis
\begin{align}
S_{Q}^{-}\left( \vec{r} \right) & = 
\psi _{k_{y}}^{\downarrow}\left( \vec{r} \right) ^{\dag }\psi _{k_{y}-Q}^{\uparrow }\left( \vec{r} 
\right)~,~ S_{Q}^{+}\left( \vec{r} \right) 
=\psi_{k_{y}-Q}^{\uparrow} \left( \vec{r} \right)^{\dag }\psi _{k_{y}}^{\downarrow }\left( \vec{r} \right) ~,
\label{eq16} \\
S_{0}^{z}(\vec{r})& = 
\psi _{k_{y}}^{\uparrow }\left( \vec{r} \right) ^{\dag }\psi _{k_{y}}^{\uparrow }\left( \vec{r} \right) - 
\psi_{k_{y}}^{\downarrow }\left( \vec{r} \right) ^{\dag }\psi_{k_{y}}^{\downarrow }\left( \vec{r} \right) ~,  \notag
\end{align}
which commute with the system Hamiltonian owing to the shifting property (\ref{eq15}) and the SU(2) symmetry of the system~\cite{bernevig}.  The non-diagonal operators
\begin{equation}
S_{Q}^{\pm }\left( \vec{r} \right) =\exp 
\left( \pm 2 i a \gamma x  \pm i b \gamma y \right) D_{\mu}\left[ x-X(k_{y})+b \gamma \right]^2 
\label{eq17}
\end{equation}
represent long-lived {\it spin edge helices} of 2DES in the presence of a perpendicular magnetic field. Here the coordinate $y$ is also dimensionless, $y \rightarrow y l_{B}/\sqrt{2}$. As seen from Eq.~\ref{eq17}, in contrast to the spin edge states, the helical edge modes have {\it finite spin-dependent phase factors} in both configurations. Going back to the initial spin basis we see that in the (a) configuration ($a=1$ and $b=0$) the factor $\exp \left( \pm 2 i\gamma x\right) $, existing also for the edge states, describes the spin precession in $(x,z)$ plane with the precession angle $\vartheta(x) =4\gamma x$, depending on the transverse $x$ coordinate, along which the helices are confined by the magnetic field via the factor $D_{\mu}\left[ x-X(k_{y})+b \gamma \right]^2$. On the contrary, in the (b) configuration ($a=0$ and $b=2$) the factor $\exp \left( \pm 2 i \gamma y\right) $ is inherent only to the SEH. This factor arises from the plane wave functions in Eq.~\ref{eq5} due to the shifting property (\ref{eq15}) and describes the spin precession in $(y,z)$ plane with the precession angle $\vartheta(x) =4\gamma y$, depending on the $y$ coordinate along the free propagation direction of the SEH.
Thus the combined effect of the perpendicular magnetic field and the confining potential on the SEH critically depends on the orientation of the edges of 2DES relative to the crystallographic axes; as we will see it depends also on the strength of the magnetic field.

In quantizing magnetic fields the spin helices are strongly localized along the transverse direction $x$ around its average position $\overline{x}_{l}\left( k_{y}\right)$.
In the limit of large negative $k_{y}$, 
$\overline{x}_{l}\left( k_{y}\right)$ is approximately independent of $k_{y}$ and takes discrete values $\overline{x}_{l}$ in the different channels $l$~\cite{smb2009,smb2001}. Therefore, in the (a) configuration the precession angle is  \textit{quantized} around $\vartheta_{l}=4\gamma\overline{x}_{l}$ for the spatially separated edge channels $l$ (see bottom of Fig.~3.a). 
Meanwhile in the (b) configuration the precession of spin by an angle $\vartheta_{l}=4\gamma y$ develops a spatially periodic structure along the $y$ direction (see bottom of Fig.~3.b), similarly to the usual PSH in an infinite 2DES in the magnetic field free case~\cite{bernevig,chang1,chang2,chang3}. Thus, by switching between the (a) and (b) configurations, i.e. by tuning the BR and D coupling strengths either to the $\alpha_R=\alpha_D$ or $\alpha_R=-\alpha_D$ case, one can realize a selection mechanism between the two alternatives of the quantized spin edge helices and of the free spin edge helices.

In the limit of large positive $k_{y}$, the electrons are localized in quasibulk Landau states around their rotation center $X\left( k_{y}\right)\gg 1 $, which increases linearly with $k_{y}$~\cite{smb2009,smb2001}. However, for a given $k_{y}$ one can neglect weak oscillations of the spin precession angle $\vartheta_{l}$  since in strong fields $l_{B}$ is much smaller than the spin-orbit length, $\lambda_{SO}=1/2m\alpha$, and the variation of $\vartheta_{l}$ during the period of the cyclotron rotation is of the order of $\gamma=2m\alpha l_{B} \ll1$. Thus, the precession angle in the configuration (a), $\vartheta_{l}\approx 4\gamma X(k_{y})$, is independent of $y$ and varies only along $x$ with $k_{y}$ (see top of Fig.~3a) while in the (b) configuration, $\vartheta_{l}\approx 4\gamma y$ is independent of $x$ but varies along $y$ (see top of Fig.~3b). 

\begin{figure}[t]
\includegraphics[width=1\linewidth]{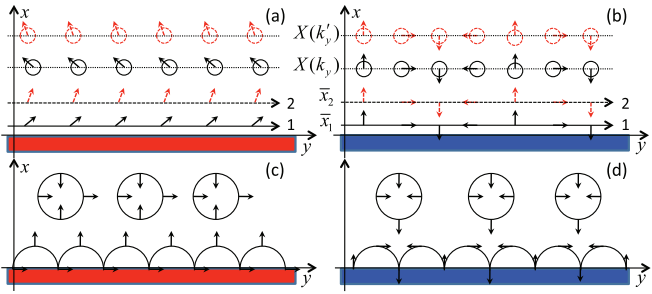}
\caption{Spin edge helices in a perpendicular magnetic field. Figs.~(a) and (c) (Figs.~(b) and (d)) correspond to the (a) configuration (the (b) configuration). Figs.~(a) and (b) (Figs.~(c) and (d)) correspond to the limit of strong (weak) magnetic fields. The bottom (top) of each figure represents the helical structures in the edge channels (in the quasibulk Landau states).}
\label{helix}
\end{figure}

In weak fields and for $l\gg1$, the quasi-classical description is valid and the $x$ coordinate oscillates in time as $x=X \left(k_{y}\right) +\sqrt{2\nu+1}\cos\omega_{B}t$. For the edge states $X\left( k_{y}\right) $ can be negative, $-\infty<X_{s}\left(k_{y}\right) \leqslant\sqrt{2\nu+1}$ so that 
in the (a) configuration the spin precesses along $x$ by an angle $0\leqslant\vartheta\leqslant2\vartheta_{0}$ where $\vartheta_{0}=4\gamma\sqrt{2\nu+1}$. Meanwhile for the bulk Landau states $X\left( k_{y}\right) \geqslant\sqrt{2\nu+1}$ and 
the precession angle in the (a) configuration varies within the range $\vartheta_{s}-\vartheta_{0}\leqslant\vartheta\leqslant \vartheta_{s}+\vartheta_{0}$. Therefore, in the magnetic fields and electron concentrations such that 
4$\gamma\sqrt{2\nu_{F}+1}=\pi/2$ with $2\nu_{F}+1=E_{F}/\hbar\omega_{B}$, the spin resonance effect takes place as shown in the top of Fig.~3c: the spin makes one full precession each time when the electron makes a full cyclotron rotation. Recently, the spin resonance phenomenon, driven by the SOI and perpendicular magnetic fields, has been experimentally demonstrated in Ref.~\onlinecite{wegscheider} by observing a strong suppression of spin relaxation. 
As seen in the top of Fig.~3d and 3c the spin resonance for the Landau states in the (b) configuration differs from that in the (a) by a simple rotation of the cyclotron orbit because in the (a) $\vartheta=4\gamma y$ depends only on $y$ while in the (b) $\vartheta=4\gamma x$ depends only on $x$.

For large negative $k_{y}$ the physical picture of the spin resonance in the (a) and (b) configurations differs essentially  ({\it cf.} the bottom of Figs.~3c and 3d). In (a) $\vartheta=4\gamma x$ depends only on the $x$, along which the electron motion oscillates within a finite range. Therefore, the precession angle is independent of the electron motion along the edge. In contrast, in (b) $\vartheta=4\gamma y$ depends on the $y$ coordinate, along which the electron propagates freely.
Therefore, in (b) the spin precession generates a spatially periodic structure of the SEH along $y$  and as seen in the bottom of Fig.~3d in the resonance case the spin changes its sign each time the electron makes a half circle in its skipping orbit.


In conclusion, we present a theory of persistent spin edge helices, which exhibit novel features, tuneable by electric and magnetic fields. We show that either a periodic structure of spin edge helices along the sample edges or a helical structure with a quantized precession angle along the transverse direction is realized. 

We thank E. Rashba, G. Vignale, and M. Glazov for useful discussions and acknowledge support from the EU Grant PIIF-GA-2009-235394 and the DFG SFB 689.

\end{document}